\documentclass{emulateapj}

\newcommand\beq{\begin{equation}}
\newcommand\eeq{\end{equation}}
\newcommand\beqar{\begin{eqnarray}}
\newcommand\eeqar{\end{eqnarray}}
 
\begin{document}

\title{Assessing the Significance of Apparent Correlations\\
Between Radio and Gamma-ray Blazar Fluxes}

\author{V. Pavlidou\altaffilmark{1,2}, J.L. Richards\altaffilmark{3}, W. Max-Moerbeck, O. G. King, T.J. Pearson, A.C.S. Readhead, R. Reeves, and M.A. Stevenson}
\affil{California Institute of Technology, Owens Valley Radio Observatory,  
Pasadena, California 91125}
\author{E. Angelakis, L. Fuhrmann, and  J.A. Zensus}
\affil{Max-Planck-Institut f\"{u}r Radioastronomie, Bonn 53121, Germany}
\author{M. Giroletti}
\affil{INAF Istituto di Radioastronomia, Bologna, Italy}
\author{A. Reimer}
\affil{Leopold-Franzes-Universit\"{a}t Innsbruck, Austria}
\author{S.~E. Healey, R.W. Romani, and M.S.~Shaw}
\affil{Department of Physics/KIPAC, Stanford University, Stanford, CA 94305, USA}
\altaffiltext{1}{Einstein Fellow}
\altaffiltext{2}{current address: Max-Planck-Institut f\"{u}r Radioastronomie, Bonn 53121, Germany}
\altaffiltext{3}{current address: Department of Physics, Purdue University, 525 Northwestern Ave, West Lafayette, Indiana 47907}

\begin{abstract}

Whether a correlation exists between the radio and gamma-ray flux densities of blazars is  a long-standing question, and one that is difficult to answer confidently because of various observational biases which may either dilute or apparently enhance any intrinsic correlation between radio and gamma-ray luminosities. We introduce a novel method of data randomization to evaluate quantitatively the effect of these biases and to assess the intrinsic significance of an apparent correlation between radio and gamma-ray flux densities of blazars. The novelty of the method lies in a combination of data randomization in luminosity space (to ensure that the randomized data are intrinsically, and not just apparently, uncorrelated) and significance assessment in flux space (to explicitly avoid Malmquist bias and automatically account for the limited dynamical range in both frequencies). 
The method is applicable even to small samples that are not selected with strict statistical criteria. For larger samples we describe a variation of the method in which the sample is split in redshift bins, and the randomization is applied in each bin individually; this variation is designed to yield the equivalent to luminosity-function sampling of the underlying population  in the limit of very large, statistically complete samples. We show that for a smaller number of redshift bins, the method yields a worse significance, and in this way it is conservative: although it may fail to confirm an existing intrinsic correlation in a small sample that cannot be split into many redshift bins, it will not  assign a stronger, artificially enhanced significance.  We demonstrate how our test performs as a function of number of sources, strength of correlation, and number of redshift bins used, and we show that while our test is robust against common-distance biases and associated false positives for uncorrelated data, it retains the power of other methods in rejecting the null hypothesis of no correlation  for correlated data. 

\end{abstract}


\maketitle 

\section{Introduction}

Whether the radio and the gamma-ray luminosities of blazars are intrinsically correlated is a long-standing debate. The presence or absence of such a correlation could provide insight into blazar emission physics. At radio frequencies low enough that synchrotron emission is self-absorbed on physical scales likely to be associated with gamma-ray emission, measurements of the gamma-ray and radio flux densities typically probe different parts of the blazar jet. If concurrently-measured, time-averaged flux densities at self-absorbed radio frequencies and high-energy ($\geq 100$ MeV) gamma-rays are intrinsically correlated, the implication would be that emission and flaring in different parts of blazar jets are driven by the same disturbances. In this case, further progress on the sequence of events that produce blazar flares can be made through high-cadence monitoring in both wavebands. If on the other hand radio and gamma-ray flux densities can be shown to be uncorrelated (a statement that needs to be carefully distinguished from the absence of evidence for correlation) then it is more likely that, over the timescales used for the flux-averaging,  emission regions probed by radio and gamma-ray observations evolve and radiate independently. 
Furthermore, should an intrinsic correlation between gamma-ray and radio flux densities be unambiguously demonstrated, radio blazar luminosity functions could be used to establish the shape and normalization of gamma-ray luminosity functions or $\log N-\log S$ distributions (however, proper care should be exercised to account for any significant scatter in the correlation, see, e.g., the discussion in Ackermann et al.~2011). From there, the unresolved blazar contribution to the diffuse gamma-ray background could be estimated (e.g., Stecker \& Salamon 1996; Kazanas \& Perlman 1997; Stecker \& Venters 2010). This is particularly important as blazars constitute a guaranteed background for any search in the diffuse gamma-ray emission for yet-undetected classes of sources such as galaxy clusters, and for signatures of exotic physics. 

 Strong correlations between radio and gamma-ray luminosities have been claimed based on EGRET data (e.g., Stecker et al.\ 1993; Padovani et al.\ 1993; Stecker \& Salamon 1996). However, these findings have been disputed (e.g., M\"{u}cke et al. 1997; Chiang \& Mukherjee 1998) based on more detailed statistical analyses. The objections against the claimed correlations can be summarized as follows. 

First, artificial flux-flux correlations can be induced due to the effect of a common distance modulation of gamma-ray and radio luminosities. Feigelson \& Berg (1983) have argued that in statistically complete surveys of relatively small depth, apparent flux-flux correlations do not appear unless the corresponding luminosities are intrinsically correlated: if luminosities are intrinsically uncorrelated most objects will only have an upper limit rather than a detection in one of the wavebands.  However this is not the case in samples that are selected with complex or subjective criteria, samples in which there is clustering around a preferred luminosity value, samples in which detection in both wavebands is one of the selection criteria, or samples in which the luminosity dynamical range is, for any reason, small compared to the distance modulation range.  In such cases, the application of a common distance-squared factor to both radio and gamma-ray luminosity will automatically induce an artificial flux-flux correlation. 

This effect cannot be avoided simply by searching for correlation in luminosity space, as the danger of inducing an artificial apparent correlation is even greater in this case due to Malmquist bias:  in flux-limited (or approximately flux-limited) surveys,  most objects are concentrated close to the survey sensitivity at each wavelength. By modulating these limiting fluxes by  a common distance factor to return to luminosity space, artificial correlations arise.

Finally,  the data used to obtain the claimed correlations were not synchronous. The direction in which non-simultaneity affects any intrinsic correlation is unclear. On the one hand, non-simultaneous data may wash out an intrinsic correlation which might otherwise be found in concurrently measured data. On the other hand, the tendency to detect more flaring objects than objects in a quiescent state in surveys may lead to enhanced correlations, essentially representing peak flux / peak flux correlations of  different flares, which although they may be indicative of the overall energetics of flares in a single object, they do not convey any detailed information regarding the time-averaged behavior of the object. 
In the {\it Fermi} era, the possibility of a correlation between gamma-ray and radio fluxes of blazars has generated a lot of interest, and the question has been explored using {\it Fermi} Large Area Telescope (LAT) fluxes in combination with archival (Ghirlanda et al.~2010, Mahony et al.~2010, Giroletti et al.~2010; Ackermann et al.~2011), quasi-concurrent (Kovalev et al.~2009) and concurrent (Giroletti et al.~2010; Ackermann et al.~2011, Angelakis et al.~2010; Fuhrmann et al.~2012, in preparation) radio data.

The intrinsic significance of an apparent correlation between radio and gamma-ray flux densities in strictly flux-limited, large datasets is relatively straight-forward to assess, by Monte-Carlo draws from the underlying luminosity functions in both datasets, obeying the same selection criteria as the observed sample of sources (e.g., Bloom 2008). In practice however we frequently encounter the case where a sample of monitored sources has been selected to optimize the likelihood of high-impact observations in individual objects using complex and often subjective criteria, which are difficult to reproduce in a simulation. Although such samples are not ideally configured for unbiased population studies, they may present significant advantages in other respects, such as multi-band coverage, high cadence of observations, and simultaneity between different waveband data. It is thus important to be able to assess as robustly as possible the intrinsic significance of any apparent correlations observed in such samples.  Here, we introduce a method for the quantitative assessment of the significance of a correlation in such cases, based on permutations of observed flux densities, while ensuring that the dynamical ranges in luminosity and flux density are kept fixed. When this method is applied in large, statistically complete samples that are split in redshift bins, it asymptotically approaches luminosity-function sampling. For smaller samples, the significances it returns are conservative: existing intrinsic correlations may not be verified, but  exaggerated significances are avoided. 
Our method has been recently used by the {\it Fermi-}LAT collaboration (Ackermann et al.~ 2011) to study the correlation between GeV and cm radio fluxes (both archival and concurrent, the latter from the Owens Valley Radio Observatory 15 GHz monitoring program, Richards et al.~2011\footnote{Program description and data also available online, at: \\{\tt http://www.astro.caltech.edu/ovroblazars/}}). They have established, at a very high significance level, the existence of a positive correlation ($<10^{-7}$ probability of the correlation arising by chance). Our method is also currently used in studies of multi-frequency concurrent radio observations by the F-GAMMA program (Angelakis et al.~2010; Fuhrmann et al.~2012, in preparation). Here, we discuss in detail the method and its implementation, and we evaluate its performance using both simulated and real ({\it Fermi} and OVRO) data. 

  We caution the reader that our proposed algorithm assumes perfectly concurrent data and thus does not address any possible effects of non-simultaneity. In addition, we stress that our method cannot compensate for sample selection effects or incompleteness relative to a parent population. For example, if the objects in the examined sample do not constitute a representative sample of the blazar population, even when a statistically significant correlation between radio and gamma-ray flux densities can be established in the objects of the observed sample, it is not possible to generalize this result to the blazar population as a whole. This limitation can only be addressed by more careful sample selection.

This paper is organized as follows. In \S 2 we discuss our method, and in \S 3 we present in detail the  implementation of the statistical test we have adopted. Demonstrations of the test and evaluations of its performance are presented in \S 4. We summarize and discuss our conclusions in \S 5.

\section{Method}

\subsection{Small, subjectively selected samples}\label{original}
The purpose of the test is to quantitatively assess the significance of an apparent correlation between concurrent radio and gamma-ray flux densities of blazars in the presence of distance effects and subjective sample selection criteria. We will do so by testing the hypothesis that emission in the two wavebands is intrinsically uncorrelated: we will calculate how frequently a sample of objects similar to the sample at hand, with intrinsically uncorrelated gamma/radio luminosities,  will yield an apparent correlation as strong as the one seen in the data, when subjected to the same distance and dynamical-range effects as our actual sample. 

In our implementation of the test, the strength of the apparent correlation is quantified by the Pearson product-moment correlation coefficient $r$ (Fisher 1944), defined as 
\begin{equation} \label{rcoeff}
r = \frac{\sum_{i=1}^N(X_i-\bar{X})(Y_i-\bar{Y})}
{\sqrt{\sum_{i=1}^N(X_i-\bar{X})^2\sum_{i=1}^N(Y_i-\bar{Y})^2}}\,,
\end{equation}
with $(X_i,Y_i)$ in our case being a pair of the  logarithms of the flux densities in each frequency for a single object. The reason for taking the logarithm is two-fold.  First, it ensures that, for sources with a power-law distribution of fluxes, there will not be a clustering of most measurements around the low-flux corner of the flux-flux plane, which would then allow single high-flux outliers to induce an artificially high $r$ value. Second, it linearizes any power-law relation between the variables, which improves the behavior of correlation measures that target specifically the linear correlation between variables (such as the Pearson $r$). 

This test can also be used with any statistic quantifying correlation strength instead of the Pearson product-moment coefficient, including non-parametric correlation measures (e.g., Siegel \& Castellan 1988; Conover 1999).

Since the sample selection criteria are assumed to be subjective, the challenge in defining our test lies in constructing simulated object samples of intrinsically uncorrelated gamma/radio flux densities, similar in other respects to our actual object sample. 
In order to overcome this difficulty, we use only permutations of measured quantities. 

Our method is a variation of a classical permutation test for the assessment of a correlation (e.g., Wall \& Jenkins 2003, \S 4.2.3; see also Efron \& Petrosian 1998 for permutation methods for doubly truncated datasets). Its novelty lies in the fact that while we are trying to establish a correlation between flux densities and calculate a distribution of correlation measures in simulated sets of flux density logarithms, we perform permutations in luminosity space (see also Fender \& Hendry 2000 for a similar Monte Carlo approach of evaluating  an apparent distance-squared effect and the possible effect of Doppler beaming in the case of radio data of persistent X-ray binaries). In this way, we can simulate the effect of a common distance on intrinsically uncorrelated luminosities,  by applying a common redshift to permuted luminosity pairs to return to flux space. By assessing the significance  in flux space we  avoid Malmquist bias, and we automatically account for the limited flux dynamical ranges in the two frequencies under consideration. 

We do so as follows:
\begin{enumerate}
\item From the measured radio and gamma-ray flux densities, we calculate radio and gamma-ray luminosities at a common rest-frame radio frequency and rest-frame gamma-ray energy. 
\item We permute the evaluated luminosities, to simulate objects with intrinsically uncorrelated radio/gamma luminosities.
\item We assign a common redshift (one of the redshifts of the objects in our sample, randomly selected) to each luminosity pair, and return to flux-density space.  Assigning a common redshift allows us to simulate the common-distance effect on uncorrelated luminosities. Using measured redshifts and luminosities guarantees that the distance and luminosity dynamical range in our simulated samples is also identical to that of our actual sample. 
\item To avoid apparent correlations induced by a single very bright or very faint object much brighter or fainter than the objects in our actual sample, we reject any flux-density pairs where one of the flux densities is outside the flux-density dynamical range in our original sample. 
\end{enumerate}

Using a randomly selected set of flux density pairs, with number equal to the number of objects in our actual sample, we calculate a value for $r$. We repeat the process a large number of times, and calculate a distribution of $r-$values for intrinsically uncorrelated flux densities. The fraction of  $|r|\geq r_{\rm data}$, where $r_{\rm data}$ is the $r-$value for the observed flux densities, is the probability to have obtained an apparent correlation at least as strong as the one seen in the data from a sample with intrinsically uncorrelated gamma-ray/radio emission. This quantifies the statistical significance of the observed correlation. 

Formally, the null hypothesis tested with this procedure is 
$H_0:$ {\em The radio and gamma-ray luminosity of blazars are independent, and redshift is independent of both luminosities.} We note that in many cases, this is not the hypothesis we would like to be testing, as luminosities depend on redshift in most population models of active galactic nuclei. Ideally, we would like to test for independence between radio and gamma-ray luminosities conditioned on redshift. However, this is not always practically possible due to sample size and redshift span of the sources. For the cases when the sample size is large enough and the sources included in the sample are adequately spread over redshifts, the test discussed in the next subsection will fulfill this requirement. For cases however when sample limitations are prohibitive for such a study, we show that testing $H_0$ {\em with the implementation presented in this work} can provide a conservative alternative to the full problem: if $H_0$ is rejected with high significance, then it is safe to assume that radio and gamma-ray luminosities are also not independent conditioned on redshift. However, if $H_0$ cannot be rejected, no conclusion can be reached for either hypothesis, as absence of evidence for a correlation is not equivalent with evidence for absence of a correlation. 

\subsection{Larger samples: splitting the sample in redshift bins}\label{zbinsdisc}

The process of pair rejection discussed in step 4 above 
may alter the distribution of luminosities, fluxes, and redshifts of the randomized data and introduce substantial differences from the corresponding distributions of the original dataset. 

The cause of this effect is the randomization of redshifts among all sources, and it is straight-forward to understand. Low-luminosity nearby objects, when combined with large redshifts, will result in very faint fluxes which are outside the original flux dynamical range and thus rejected. For this reason, the simulated datasets will have fewer very-low--luminosity objects compared to the original dataset. In addition, rare, high-luminosity, high-redshift objects, when combined with low redshifts, will result in very high fluxes, also outside the original flux dynamical range and thus rejected. For this reason, the simulated datasets will also have fewer very-high--luminosity objects compared to the  original dataset. In contrast, the number of intermediate-luminosity objects will be relatively enhanced in simulated datasets. The distributions of redshifts and fluxes of the simulated datasets will also be altered for similar reasons. 

If the pair rejection rate is high, the properties of the simulated datasets could be different than the properties of the original dataset, and these biases could affect our estimation of a correlation significance. In small and subjectively selected datasets, this problem is a necessary evil. The effect of these biases is,  as we will show below, to worsen the estimated significance of a correlation, rather than induce false positives of enhanced significance.  However, in the case of larger samples, there is a simple alteration in the methodology described in \S \ref{original} that can significantly alleviate these biases: splitting the sample in redshift bins. 
 
In this variation of the test, the original sample is split into a number of bins dependent on the available number of objects (as we discuss below, we need about 10 objects or more per bin, and in any case no fewer than 8). We then generate randomized flux-density pairs in each redshift bin with the process described above. Because the range of redshifts that are permuted between objects of different luminosities is much smaller, the likelihood that one of the resulting randomized flux densities will exceed the flux-density dynamical range of the original dataset is much smaller. As a result, the pair rejection rate is decreased, and the luminosity, redshift, and flux distributions of the randomized data pairs resemble more closely those of the original dataset. 

The similarity between distributions of the randomized and the original data increases as the size of the sample increases and the width of each redshift bin decreases.  If the original dataset is also a  statistically complete and flux-limited sample, then the test asymptotically approaches the luminosity-function--sampling test as the size of the original dataset approaches infinity and the size of each redshift bin used approaches zero. This can be understood as follows. In the limit of zero-size redshift bin, all objects within a single redshift bin are at the same distance. Therefore, permuting the luminosities of objects at that distance is equivalent to forming luminosity pairs by randomly sampling each frequency's luminosity function at a specific redshift and with a specific flux-density limit (the limit of the original sample). Repeating the process at all redshift bins is equivalent to sampling the luminosity functions at all redshifts. Then, the ``pool'' of randomized data pairs, from which we draw the mock datasets, could have been equivalently generated through luminosity function sampling. 

Formally, the null hypothesis tested with this procedure is 
$H_0:$ {\em Conditional on redshift, the radio and gamma-ray luminosity of blazars are independent,} which is the hypothesis that one would generally wish to test. For this reason, this version of the test should be preferred whenever possible. 

\section{Implementation}

In this section we describe how  the method discussed above can be implemented in practice for small and large datasets. 

\subsection{Small, subjectively selected samples}\label{impsimple}

The first step is to convert the blazar gamma-ray fluxes (which are usually reported as integrated photon fluxes $F$ above some fiducial energy $E_0$, usually 100 MeV), to energy flux densities, so that the comparison with radio flux densities can be done on an equal footing\footnote{Other possible choices is to correlate radio flux densities with gamma-ray photon fluxes at some particular energy bin, or with the integrated photon fluxes themselves (see, for example, Abdo et al.~2011). In these cases, Eq. \ref{mockg} should be changed accordingly.}. We do so by assuming that the photon fluxes are power laws, so that the flux (number of photons per unit area-time-energy bin) is
\beq
\frac{dN_{\rm photon}}{dE\, dA\, dt} = F_0\left(\frac{E}{E_0}\right)^{-\Gamma}\,.
\eeq
In this case, the gamma-ray energy flux density $S_\gamma \equiv dE/dE\, dA\, dt $ at $E_0$ is given by $S_\gamma(E_0) =F_0E_0 = F(\Gamma-1)$ and its energy dependence is 
\beq\label{fdensity}
S_\gamma (E) = (\Gamma-1)F\left(\frac{E}{E_0} \right)^{-\Gamma+1}\,.
\eeq

The relation between monochromatic flux density $S(\nu)$ and monochromatic luminosity $L(\nu)$ for a source at redshift $z$ is 
\begin{equation}\label{basic}
S(\nu)= \frac{L[\nu(1+z)]}{4\pi d^2(1+z)}
\end{equation}
where $d=(c/H_0)\int_0^z dz/\sqrt{\Omega_\Lambda+\Omega_m(1+z)^3}$. Here $H_0$ is the present-day value of the Hubble parameter, and $\Omega_\Lambda$ and $\Omega_m$ are the vacuum energy and matter density parameters. In this work, we have used $\Omega_m = 0.26$ and $\Omega_\Lambda = 1-\Omega_m$, consistent with, e.g., Larson et al. (2011). Note that the value of $H_0$ drops out of the calculation as $d$ in the formalism we describe below appears only in ratios.   
If the source has a spectral index $\alpha$ so that $S(\nu) \propto \nu^\alpha$ at the frequency of interest, Eq. (\ref{basic}) implies that the relation between $S(\nu)$ at observer-frame $\nu$ and $L(\nu)$ at rest-frame $\nu$  (the K-correction) is
\beq\label{kc}
L(\nu) = S(\nu)4\pi d^2(1+z)^{1-\alpha}.
\eeq
So if a radio flux density  $S_r(\nu)$ (at observer-frame $\nu$)  is turned into a luminosity density (at rest-frame $\nu$) using a redshift $z_1$ and a spectral index $\alpha_r$,  and this luminosity density is then returned to flux-density--space (at observer-frame $\nu$)  using a different redshift $z'$ but the same spectral index $\alpha_r$,  we can write
\beq \label{mockr}
S'_r(\nu) = S_r(\nu) \left(\frac{d_1}{d'}\right)^2
\left(\frac{1+z_1}{1+z'}\right)^{1-\alpha_r}\,,
\eeq
where $d_1=d(z_1)$ and $d'=d(z')$. 
For the same procedure with gamma-ray flux densities and a source at a redshift $z_2$ we can write
\beq \label{mockg}
S'_\gamma(E_0) = (\Gamma-1)F\left(\frac{d_2}{d'}\right)^2
\left(\frac{1+z_2}{1+z'}\right)^\Gamma\,.
\eeq
In practice, we perform the following steps. 
\begin{itemize}
\item[(i)] For each blazar, we use the flux density in radio and gamma-ray frequency to produce monochromatic luminosities at the same (now rest-frame) frequency in the two bands.
\item[(ii)] We construct all possible pairings (excluding the original ones) of radio and gamma-ray luminosities from our observed sample.
\item[(iii)] We assign a common redshift $z'$ to each permuted pair (one of the available redshifts in our sample).
\item[(iv)] We calculate ``mock'' radio and gamma-ray flux densities $S'_r, S'_\gamma$ for each pair using Eq.~(\ref{kc})\footnote{Equivalently, we can use directly Eqs.~(\ref{mockr}) and (\ref{mockg}), without explicitly calculating luminosities first.}. 
\item[(v)] We accept the pair if both flux densities are within our original flux-density dynamical range in each band, or reject it otherwise.
\item[(vi)] We randomly select $N$ pairs out of all the possible combinations, where $N$ is equal to the number of our original observations. Each set of $N$ pairs is now a simulated dataset of intrinsically uncorrelated  flux/flux observations.
\item[(vii)] For each simulated dataset, we compute $r$ using Eq. (\ref{rcoeff}), where $X_i = \log (S'_{r,i})$ and $Y_i = \log (S'_{\gamma,i})$, with $i$ running from 1 to N.
\item[(viii)] We repeat steps (vi-vii) $m$ times, where $m$ is a sufficiently large number to sample the underlying $|r|$ distribution. In our tests below $m$ is between $10^6 - 10^7$.
\item[(ix)] We calculate the probability for the observed $|r|$ to have occurred through uncorrelated flux densities from the $|r|-$values obtained in step (viii). 
\end{itemize}

Our technique can be applied to samples that are very small and still yield a reliable estimate of the distribution of $|r|$. The total number of simulated pairs that we can construct through our permutation technique from $N$ objects is $N_{\rm pairs}=N^2(N-1)$ (where we permute both flux densities as well as redshifts). Only a fraction $N_{\rm surv}$ will survive the low- and high- flux-density cuts that ensure that the flux-density dynamical range remains the same as in the original sample. Assuming a reduction no larger than a factor of 5 (i.e.~$N_{\rm surv} \gtrsim N_{\rm pairs}/5$, shown in practice to be a conservative assumption), the total number of combinations of $N$ pairings different from each other by one or more pairs out of a population of $N_{\rm surv}$ objects then is 
\begin{equation}
{\rm pair \,\,\, combinations} = \frac{N_{\rm surv}!}{N!(N_{\rm surv}-N)!} \,,
\end{equation}
which is $\gg 10^7$ for samples with $N \gtrsim 8$. However, in small datasets a statistically significant correlation is harder to establish, even if the distribution of $|r|$ can be estimated with sufficient statistics. In addition, as we will also show in \S \ref{demonstration}, the biases in the luminosity, redshift, and flux distributions of the simulated datasets introduced due to pair rejections (see discussion in \S \ref{zbinsdisc}) tend to worsen the significance that can be established through this test. 

\subsection{Splitting larger samples in redshift bins}\label{thesplit}

Whenever the size of the source sample is large enough to allow splitting in more than one redshift bins, this variation of the test is recommended, as the effect of biases introduced through pair rejection decreases with increasing number of redshift bins (decreasing redshift bin size). 

To implement this variation of the test, we split the sample in $N_z$ redshift bins. Our choice for the test implementation is to use variable redshift bin size, selected in such a way that the number of sources in each bin is as close to equal as possible, but never fewer than 8. However, other choices are also possible (for example, keeping the redshift bin size approximately equal; or splitting by luminosity distance rather than redshift, and keeping the luminosity distance bin size approximately equal).

For the sources in each one of the $N_z$ bins, we apply steps (i)-(v) of \S \ref{impsimple}. We then combine all accepted simulated data pairs from all redshift bins to generate the ``pool'' of all possible pair combinations. Finally, we apply steps (vi) - (ix) to this combined randomized pair ``pool''.

\section{Demonstrations of the test}\label{demonstration}

In this section we present example applications of our tests, using both real and simulated data, to evaluate the performance of our proposed test and demonstrate several aspects of its implementation. 
For the applications on real data, we will use gamma-ray flux measurements from {\it Fermi} LAT and radio flux-density measurements from the OVRO  40 M Monitoring Program (Richards et al.\ 2011). 
In addition, we will use simulated data to evaluate the the performance of the method: its effectiveness in rejecting false positives due to common-distance biases in correlation assessments, and its power in establishing significant correlations when such correlations do exist. As a benchmark we will use the face-value estimate of the significance for the Pearson correlation coefficient $r$, which evaluates the probability of a certain (or bigger) value of $r$ to occur by chance in the ``dart-throwing'' scenario (i.e., when pairs are randomly drawn from uncorrelated Gaussian distributions, assuming that no biases exist). In the latter scenario,  the significance only depends on the value of $r$ and the sample size $N$. In the null hypothesis (uncorrelated data), the variable
\begin{equation} \label{student}
t = \frac{r\sqrt{N-2}}{1-r^2}
\end{equation}
follows a Student's t-distribution with $N-2$ degrees of freedom. Using Eq.~(\ref{student}) significances (p-values) can be estimated for any given values of $r$ and $N$ by taking the two-tail integral of the appropriate t-distribution. 
In general, the variation of the test {\em with} redshift binning is the one which we recommend whenever possible (whenever  sample restrictions allow its use), and it is the one which we have used in our simulated datasets. 

\subsection{Demonstrations on real data}

\subsubsection{Small sample, no redshift bin splitting}\label{demo-small}

As an example of a relatively small dataset, we use the set of blazars that are included both in the LAT bright AGN source list (Abdo et al.~2009, produced using three months of LAT observations), as well as in the ``complete sample'' of the OVRO 40 M Monitoring Program (Richards et al.\ 2011). 
The latter consists of the 1158 sources north of $-20^\circ$  declination in
the Candidate Gamma-Ray Blazar Survey (CGRaBS) sample, which is a sample
of 1625 sources, mostly blazars, selected by their flux and spectral index in radio, and flux in X-rays,  to resemble the blazars detected by EGRET (Healey et al. 2008). The 1158 of the ``complete sample'' are observed approximately twice a week at 15 GHz with the  Owens Valley Radio Observatory (OVRO) 40 M Telescope.
For this study, we only use sources with known redshifts, and for which a sufficient number of high-quality 15 GHz observations were taken in the same three-month time interval of LAT observations so as to produce a meaningful concurrent 15 GHz average flux density (see Richards et al.\ 2011). This sample contains 38 sources. 

\begin{figure}
\includegraphics[width = 3in, clip]{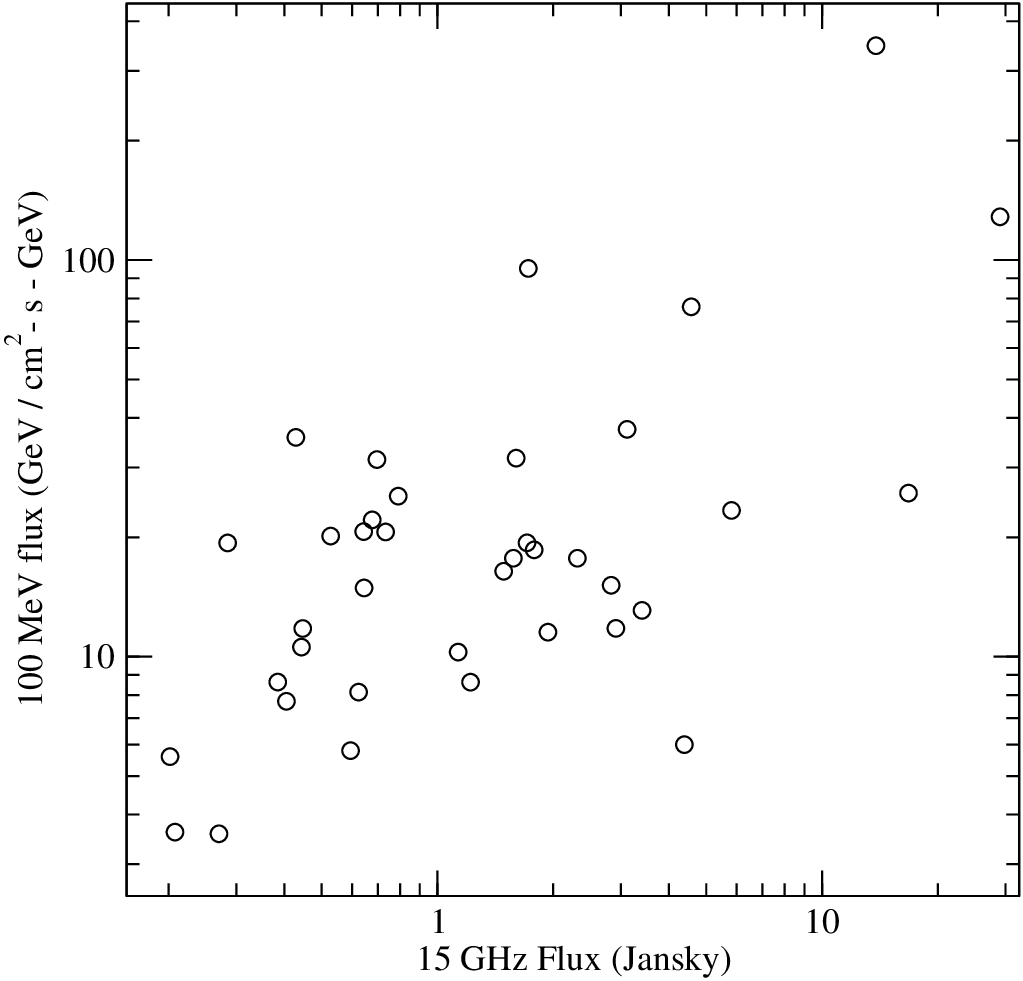}
\caption{3-month averaged concurrent 15 GHz versus 100 MeV observer-frame flux densities for the 38 blazars in our sample. \label{fluxflux}}
\end{figure}

\begin{figure}
\includegraphics[width = 3in, clip]{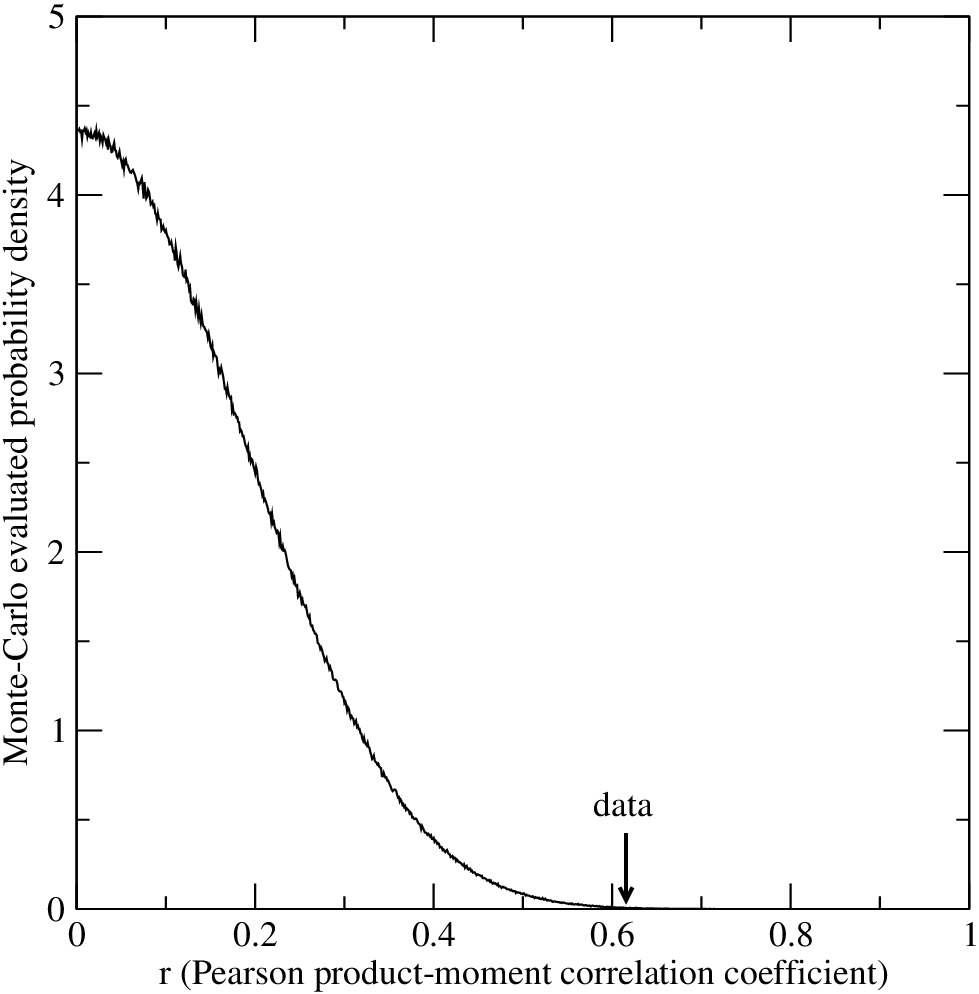}
\caption{Distribution of $|r|-$values for 38 blazars of the same dynamical range in redshift and radio and gamma-ray flux densities and luminosities as blazars in our sample. The vertical arrow indicates the $r-$value for the actual observations ($r=0.62$). The significance of the correlation is $1.5\times10^{-4}.$\label{rdistfig} }
\end{figure}

Figure \ref{fluxflux} shows 3-month averaged 15 GHz flux densities plotted against 100 MeV observer-frame flux densities obtained by integration over the same time interval for the 38 blazars in our sample. The error bars in this plot are substantially smaller than the scatter of points (see, e.g., Ackermann et al.~2011) and have been omitted for clarity. An apparent correlation between the radio and gamma-ray time-averaged flux densities is obvious, however the statistical significance of an intrinsic correlation between the radio and gamma-ray emission of these objects needs to be quantitatively assessed. To this end, we apply the data randomization analysis we have introduced in \S \ref{impsimple}.  
\begin{figure}
\includegraphics[width = 3in, clip]{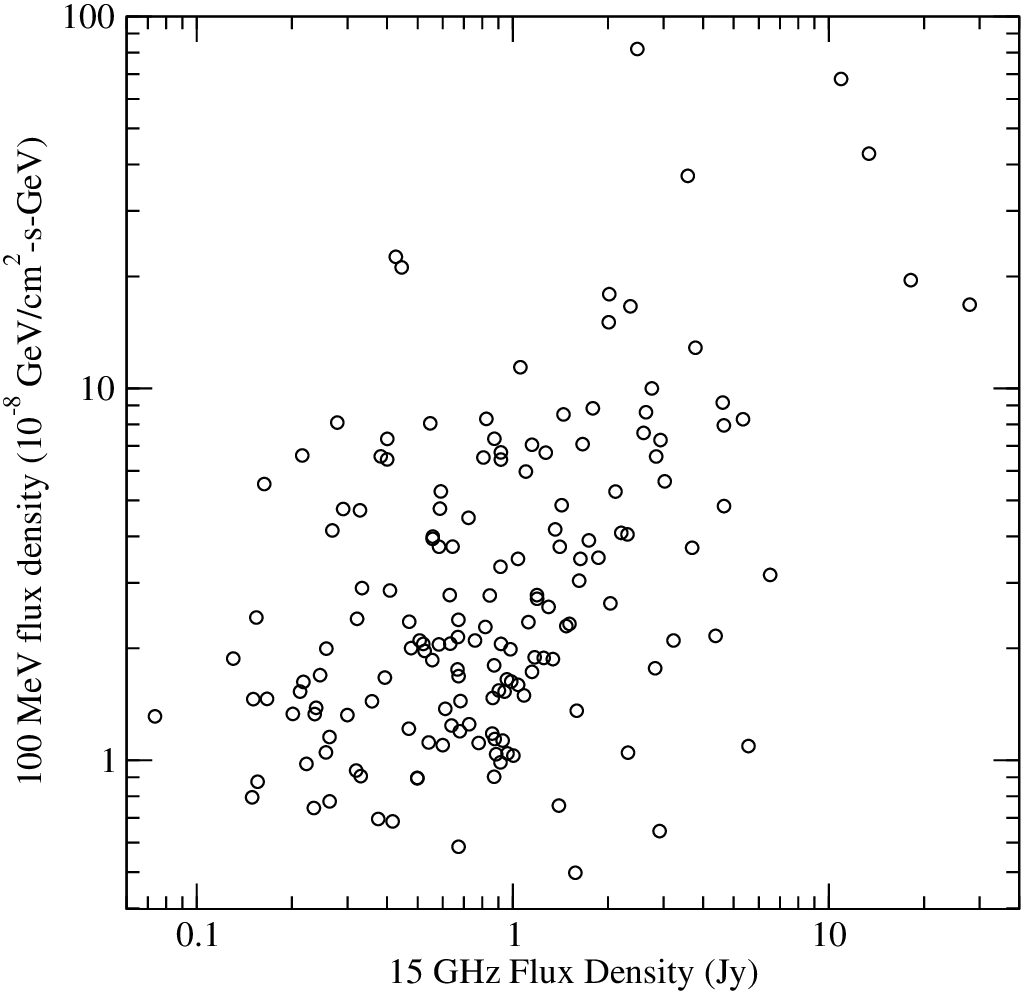}
\caption{11-month averaged concurrent 15 GHz versus 100 MeV observer-frame flux densities for the 160 blazars in the larger sample. \label{fluxfluxl}}
\end{figure}
The probability distribution of the values of $|r|$ in our simulated samples with intrinsically uncorrelated radio/gamma luminosities is shown in Fig. \ref{rdistfig}. The vertical arrow in this figure indicates the $r-$value for the observed data, equal to $0.62$.  From the 38 objects in our sample a total number of $38^2\times 37 = 53,428$ permuted pairs were generated. Of those, $13,003$ pairs had both gamma-ray and radio flux densities within the dynamical range of the original dataset.
The accepted pairs were used (in $10^7$ randomly drawn sets of 38) to generate the distribution shown in Fig. \ref{rdistfig}.  The probability to obtain  $|r|\geq 0.62$ from intrinsically uncorrelated flux-density measurements due to the effect of a common distance is $1.5\times 10^{-4}$. For comparison, the  significance estimate ignoring any biases and using only Eq.~(\ref{student})  is  $3.3\times 10^{-5}$: without a careful analysis, we would evaluate the observed correlation as more significant than we do when accounting for common-distance and flux biases, as these effects are likely to contribute at least part of the observed correlation strength. We will elaborate on the origin and quantitative behavior of this discrepancy in the following sections. 

 Note that the pair rejection rate is high - only 24\% of the permuted pairs are within the original flux density dynamical range and were accepted; biases introduced in the luminosity, flux, and redshift distributions of the simulated data are therefore a concern. However, as we will show below, were these biases absent, the significance of the correlation would improve. 

\subsubsection{Larger sample, behavior of test with increasing number of redshift bins}\label{zbinsapp}

We now turn to a demonstration of the second variation of our test, where the sample is split in redshift bins, and we discuss the alleviation of biases induced through pair rejection, and the improvement of the correlation significance with increasing number of bins.

To allow splitting in enough redshift bins to adequately demonstrate the behavior of the test in the many-bins limit we use the significantly larger sample of 160 sources that: (a) are included in the first year LAT catalog (Abdo et al.~2010); (b) are part of the OVRO 40 M telescope monitored sample; (c) have known redshifts. This same sample has been examined in detail for intrinsic correlations between 15 GHz flux density and LAT gamma-ray fluxes at various energy ranges by Ackermann et al.~(2011), using the test discussed here\footnote{Here, we use for the gamma-ray band data the 100 MeV flux density calculated according to Eq. \ref{fdensity} from integrated photon fluxes for $E>100$ MeV and using the photon index provided in 1LAC, which is different that any of the flux densities or integrated fluxes examined by Ackermann et al.~2011; this is the origin of the small differences in the value of $r$ obtained here for the data.}.

Figure \ref{fluxfluxl} shows  11-month--averaged 15 GHz flux densities plotted against 100 MeV observer-frame flux densities obtained by integration over the same time interval for the 160 blazars in the sample described above. The error bars in this plot are again substantially smaller than the scatter of points (see, e.g., Ackermann et al.~2011) and have been omitted for clarity. Through visual inspection, this sample also appears to feature an apparent correlation between radio and gamma-ray flux densities, with scatter comparable to that of the smaller sample of \S \ref{demo-small}. The correlation coefficient of the data in this case is $r=0.48$.

\begin{figure}
\includegraphics[width = 3in, clip]{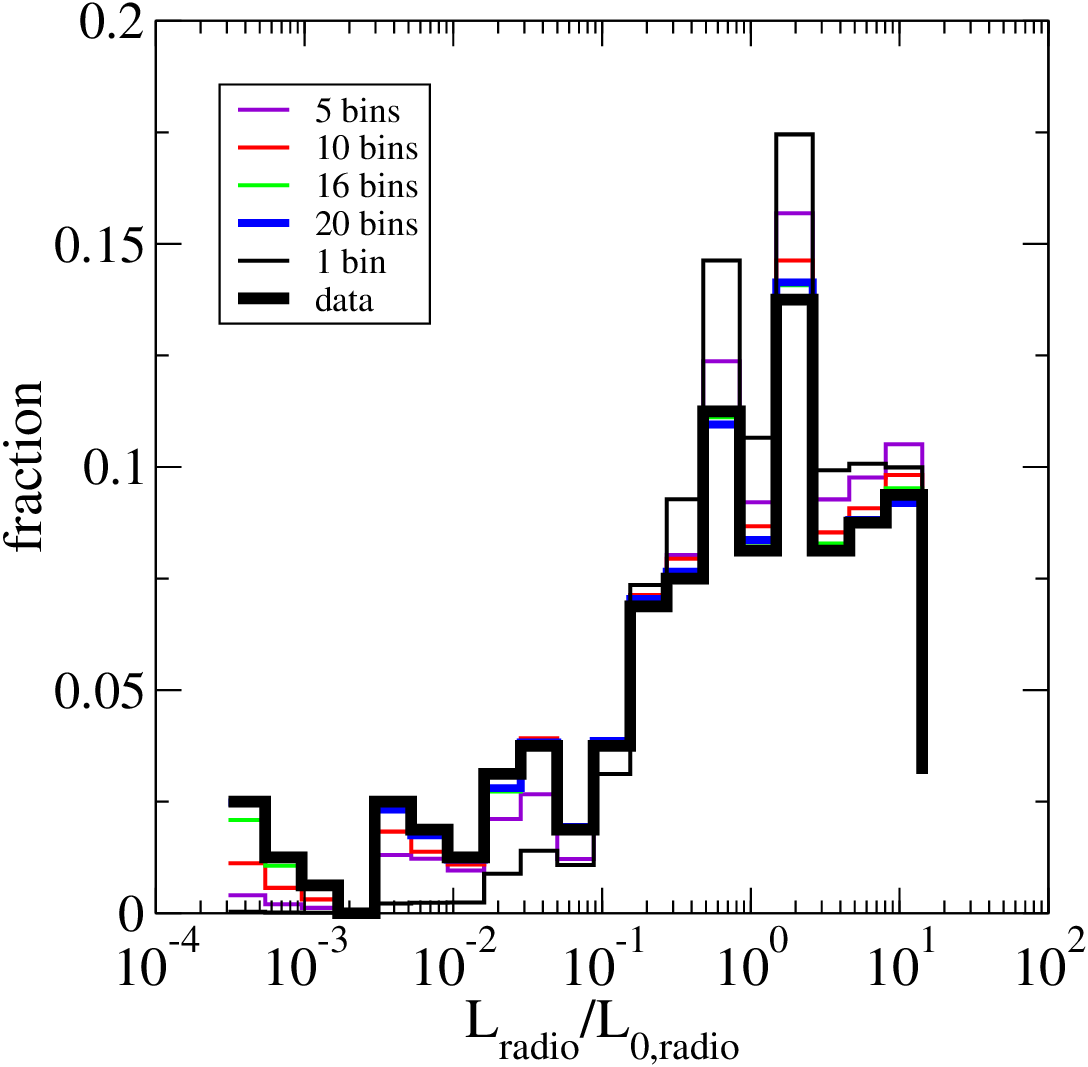}
\caption{Fraction of objects in each logarithm-in radio luminosity bin for the data (thick black line) and the accepted scrambled pairs (thin lines; different colors correspond to different numbers of redshift bins, as in legend). The radio luminosities are in units of  $L_{0,\rm radio} = 4\pi(c/H_0)^2S_0$ where $S_0=1$Jy.\label{rlumhist}}
\end{figure}

\begin{figure}
\includegraphics[width = 3in, clip]{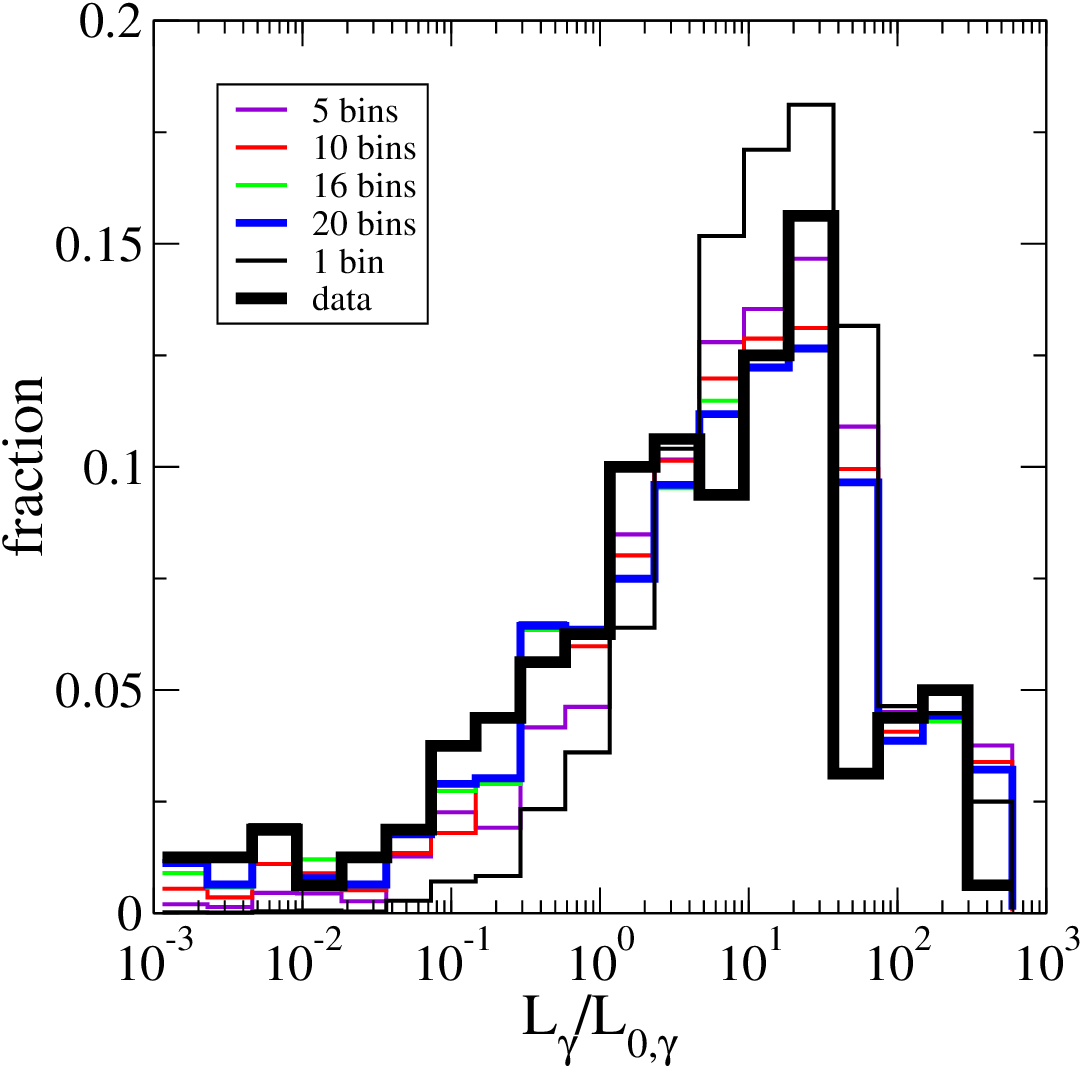}
\caption{Fraction of objects in each logarithm-in gamma-ray luminosity bin for the data (thick black line) and the accepted scrambled pairs (thin lines; different colors correspond to different numbers of redshift bins, as in legend). The gamma-ray luminosities are in units of  $L_{0,\rm \gamma}= 4\pi(c/H_0)^2S_0$ where $S_0=10^{-8} {\rm GeV/s-cm^2-GeV}$.\label{glumhist}}
\end{figure}

\begin{figure}
\includegraphics[width = 3in, clip]{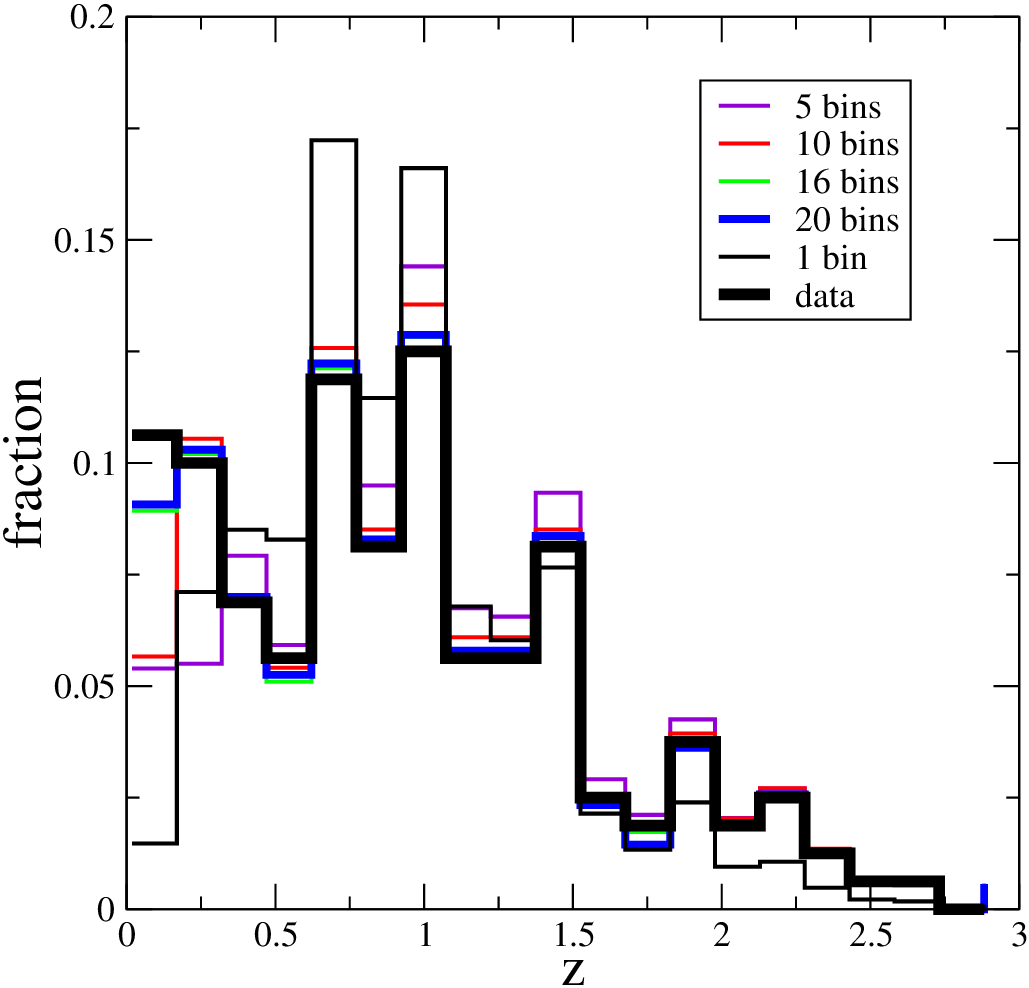}
\caption{Fraction of objects in each redshift bin for the data (thick black line) and the accepted scrambled pairs (thin lines; different colors correspond to different numbers of redshift bins, as in legend). \label{zhist}}
\end{figure}

The biases introduced through pair rejection in our first variation of the test (where the sample is not split in redshift bins) are demonstrated in Figs.~\ref{rlumhist}-\ref{zhist}. The luminosities in these figures are in units of $4\pi(c/H_0)^2S_0$, where $H_0$ is the Hubble parameter, and $S_0 = 1$ Jy for 15 GHz source-frame luminosities and $S_0=10^{-8} {\rm GeV/s-cm^2-GeV}$ for 100 MeV source-frame luminosities.

 Figure \ref{rlumhist} shows the fraction of objects in each logarithmic radio luminosity bin for the data (thick black line) and the accepted scrambled pairs (thin lines). Different line colors correspond to different numbers of redshift bins, as in the figure legend. When only one redshift bin is used (thin black line, equivalent to the first variation of our test), the shape of the luminosity distribution of the accepted scrambled pairs has a qualitatively different shape than that of the data: objects in the bins corresponding to the $\sim$3 lowest orders of magnitude in luminosity are significantly underrepresented compared to the original sample, because these low luminosities, corresponding to nearby objects in the data, are frequently rejected when they are combined with high redshifts and produce very low simulated flux densities outside the original flux density dynamical range.  When we split the sample in a larger number of bins the effect is alleviated. At 16 redshift bins the radio luminosity distribution of simulated data is very close to that of the original data, and it is essentially converged, as it does not change appreciably when the number of redshift bins is increased to 20.  

A very similar behavior for the gamma-ray luminosity distribution is shown in Fig.~\ref{glumhist}. 
In the case of the redshift distribution, shown in Fig.~\ref{zhist}, both the very low and the very high redshift bins are underestimated when no data splitting is applied (thin black line). However, at 16 redshift bins the real and simulated data distributions are very similar, and the simulated data distribution is, again, converged. In all distributions, as the number of redshift bins increases, the difference between data and simulated distributions decreases, as a result of the decreasing pair rejection rate which, at 16 redshift bins, is $\lesssim 20\%$ for all bins. 

\begin{figure}
\includegraphics[width = 3in, clip]{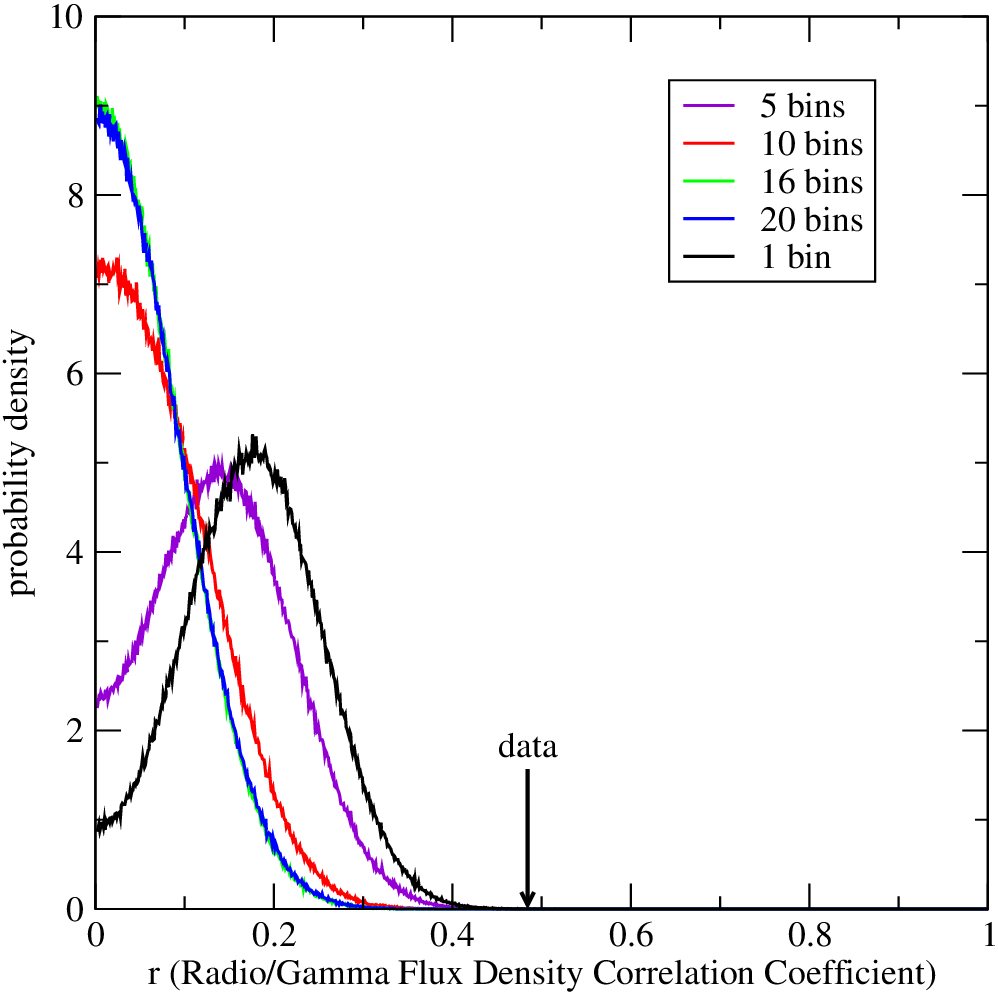}
\caption{Distribution of  $|r|-$values for randomly selected 160-blazar sets picked from the ensemble of accepted pairs generated through data scrambling. The vertical arrow indicates the $r-$value ($r=0.48$) for the actual observations. Different colors correspond to different numbers of redshift bins, as in legend. \label{multihist}}
\end{figure}
The behavior of the estimated significance as a function of the number of redshift bins is shown in Fig.~\ref{multihist}, where we have plotted the distribution of the absolute values of the correlation coefficients $|r|$ for each test implementation. Again, different colors correspond to different numbers of redshift bins used as in Figs.~\ref{rlumhist}-\ref{zhist}.  $10^6$ simulations were used to produce each curve. The $r-$value for the data is shown with the arrow.  The significance of the correlation as evaluated with 16 redshift bins is $<10^{-6}$ (if we fit the distribution shown with the blue line in Fig.~\ref{multihist} with a Gaussian, we obtain a significance of $\sim 10^{-7}$).

Again, using Eq.~(\ref{student}) to compare with the simple significance estimate based only on $r$ and $N$, we find that $t=6.88$, for which the two-tailed t-distribution with 158 degrees of freedom yields a much stronger significance of $1.3\times 10^{-10}$. The reason for this substantial difference can  be immediately understood qualitatively through inspection of Figs.~\ref{rlumhist} -\ref{zhist}. Both the radio and the gamma-ray luminosity distributions of the data show broad peaks, which means that even if there was no intrinsic correlation between radio and gamma-ray emission and radio and gamma-ray luminosities were simply randomly drawn from these distributions, values of radio and gamma-ray luminosity around the peaks would appear frequently. As a result, pairs of radio/gamma-ray fluxes corresponding to underlying luminosities clustered around likely values would be common. Such pairs, when modulated with a common distance factor, would yield an apparent correlation in flux-flux space by chance, much more frequently than if there was no peak in the luminosity distributions. The unsophisticated significance estimate contains no information about common-distance effects and the behavior of the underlying luminosity distributions, and for this reason {\bf overestimates} the significance of the apparent correlation.  However, even when these effects are accounted for using our method, the data show significant intrinsic correlation between radio and gamma-ray fluxes.

 We can see that the significance of the correlation monotonically improves with increasing number of bins\footnote{The exact statement is that the significance monotonically improves with decreasing fraction of rejected pairs. Should a particular choice in redshift binning result in increased rejection fraction, the significance would worsen, even if the number of bins was larger.}. The reason for this behavior can be understood from Figs.~\ref{rlumhist} and \ref{glumhist}. The more frequent rejection of pairs at the edges of the luminosity distributions results in the artificial enhancement of the peaks in the luminosity distributions at intermediate luminosities. This stronger peak results in an enhanced incidence of artificial correlations. As a result, the significance of the apparent correlation of the data drops.

This is also the reason for the appearance of a peak at positive values of $|r|$ in the $|r|$ distribution of simulated datasets in Fig.~\ref{multihist} when the number of redshift bins is low. However, it is not guaranteed that a small number of redshift bins will generate such a peak at positive $|r|$ - this depends on the details of the luminosity distribution of the original dataset and of the pair rejection. For example, such a peak does not appear in our smaller dataset example in Fig.~\ref{rdistfig}. Conversely, a large number of redshift bins does not guarantee that such a peak will not appear. If our original dataset is selected in such a way that a certain narrow range of luminosities is over-represented, then such a peak is intrinsic to the dataset and it will appear regardless of number of bins used. 

It is also interesting to consider the behavior of the test in the limit of a very large number of redshift bins that could be in principle used if we had a very large sample available for study, and in the case that our sample was a statistically complete, flux-limited set of sources. In this case, each redshift bin could be made very narrow, and all sources within the bin would be located essentially at the same distance. The set of radio luminosities within each bin would then be a representation of the radio luminosity function at a fixed redshift, with a limiting luminosity set by the limiting flux and the bin redshift. The set of gamma-ray luminosities within the same bin would similarly be a representation of the gamma-ray luminosity function. Since all sources would be located at the same distance, data randomization within the bin would never produce fluxes outside the original dynamical range, and no pairs would be rejected. The simulated pairs would then have exactly the same luminosity distribution as the data, and they would continue to be a representation of the luminosity functions at the two frequencies under consideration, as ``fair'' as the original data. As a result, in the limit of the ``perfect sample'' and a large number of redshift bins, our test would yield exactly the same result as a statistical test sampling random radio and gamma-ray fluxes from known luminosity functions. Our test deviates increasingly from this result as the statistical properties and the size of the sample deteriorate. 

As shown in Fig.~\ref{multihist}, our proposed test is conservative: a smaller number of redshift bins will generally result in an increased rate of pair rejection and a worse correlation significance. In this way, it is possible that a real, intrinsic correlation cannot be confirmed by this test if a poor sample is used. However, the test will not yield artificially enhanced significances. 

\subsection{Demonstrations on simulated data: \\ Uncorrelated datasets}

In this section we discuss the performance of the test when applied to datasets drawn from intrinsically uncorrelated populations. In particular, we evaluate the effectiveness of our test in rejecting false positives that we might have obtained due to common-distance biases had we used the estimate of the significance given by Eq.~(\ref{student}).  In \S \ref{dataproduction} we describe how we generate the intrinsically uncorrelated simulated datasets that we use to test the performance of our method, and in \S \ref{robustness} we examine this performance and the robustness of the evaluated significances against common-distance biases.

\subsubsection{Generation of uncorrelated simulated datasets}\label{dataproduction}
To test the performance of our method in the case of intrinsically uncorrelated data, we produce simulated datasets in the following way. 
\begin{itemize}
\item We draw a gamma-ray luminosity from a log-normal distribution, with probability density function
\begin{equation}
p(L_\gamma) = \frac{1}{L_\gamma\sqrt{2\pi\sigma_1^2}}
\exp\left[-\frac{\left(\ln L_\gamma -\mu_1\right)^2}{2\sigma_1^2}\right]\,.
\end{equation}
\item We draw a radio luminosity from a log-normal distribution, with probability density function
\begin{equation}
p(L_r) = \frac{1}{L_r\sqrt{2\pi\sigma_2^2}}
\exp\left[-\frac{\left(\ln L_r -\mu_2\right)^2}{2\sigma_2^2}\right]\,.
\end{equation}
\item For this pair, we also draw a common redshift from a uniform distribution with lower limit $z_{\rm low}$ and upper limit $z_{\rm up}$.
\item We evaluate the resulting gamma-ray and radio fluxes, and check whether they reside within an allowed flux dynamical range of three orders of magnitude. If either one does not, we reject the pair and repeat the draw. 
\item We repeat the process above until we have 30 pairs within our desired flux dynamical range. This then is our simulated, intrinsically uncorrelated dataset, to which both a common distance factor and a limit in the flux dynamical range have been applied. 
\end{itemize}

We anticipate that the effect of the common-distance biases will increase as the luminosity dynamical range decreases and the redshift dynamical range increases. This can be easily understood by considering the extreme limits. Datasets drawn from luminosity delta-functions will always appear perfectly correlated within errors: the spread in fluxes in each waveband is only due to the distance factor, which is the same in each pair, and errors. Conversely, if all sources are at the same redshift, there will be no common-distance effect: the distance factor is always the same, and any observed correlation has to be intrinsic. 

To assess when common-distance biases become important, we will use the {\em coefficient of variation} (eg., Frank \& Althoen 1995) of the redshift and luminosity distributions (standard deviation in units of the mean, $c_z$ and $c_L$ respectively) to quantify the dynamical range of each distribution. As we will see in the next section, the importance of common-distance biases is generally dependent on the ratio of the luminosity coefficient of variation to the redshift coefficient variation, $c_L/c_z$, and decreases as this ratio increases. In our simulated datasets we have used radio and gamma-ray luminosity distributions\footnote{Since the flux/flux correlation coefficient is evaluated in logarithmic space, changing the units of the luminosity, or, equivalently, the mean of the luminosity distribution, will only uniformly slide the points along the flux axes and will not affect the apparent correlation strength, as long as the flux limits are also shifted accordingly.} with $\mu_1=\mu_2 = \mu_0$ and $\sigma_1=\sigma_2=\sigma_0$ and, as a result, the same value of $c_L$, but in practice the relevant value of $c_L$ is the one of the more extended of the two distributions. For the distributions we have used here, 
\begin{equation}
c_L = \left[\exp\left(\sigma_0^2\right)-1\right]^{1/2}
\end{equation}
and 
\begin{equation}
c_z = \frac{z_{\rm up}-z_{\rm low}}{\sqrt{3}(z_{\rm up}+z_{\rm low})}\,.
\end{equation}

\subsubsection{Robustness of the test against common-distance biases}\label{robustness}

To evaluate the robustness of our test against common-distance biases and its ability to reject false positives, we generate, using the procedure described in \S \ref{dataproduction}, simulated datasets with varying values of the ratio $c_L/c_z$ of 30 objects each, and we calculate the significance of the apparent correlation using our method and the simple estimate of Eq.~(\ref{student}).

\begin{figure}
\includegraphics[width = 3in, clip]{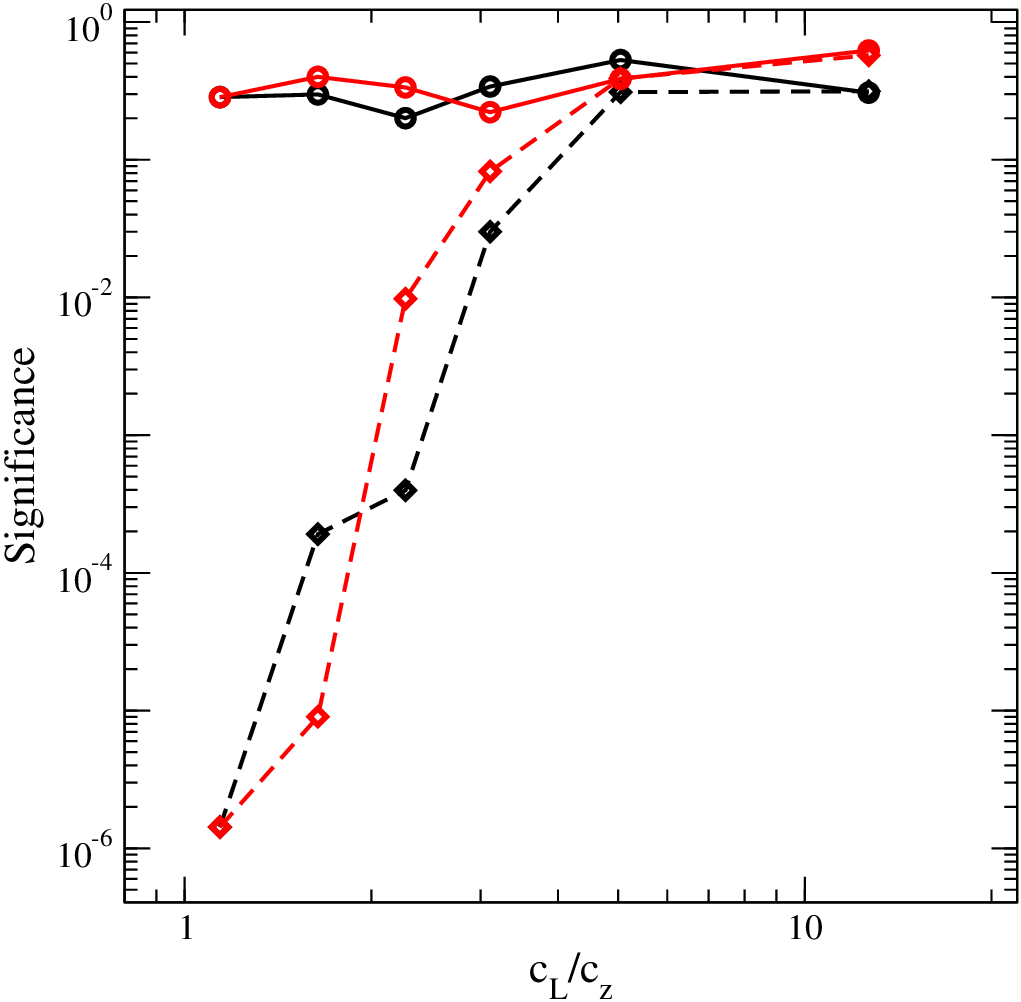}
\caption{Significance (probability to obtain an $r$ as big or bigger than the data by chance) returned by our method (circles, solid lines) compared to significance returned by the simple estimate of Eq.~(\ref{student}) which does not account for common distance biases, as a function of the ratio of coefficients of variation of the luminosity and redshift distributions. Black points were generated by varying the width of the luminosity distribution while keeping the redshift distribution fixed. Red points were generated  by varying the width of the redshift distribution while keeping the luminosity distribution fixed. Our method always succeeds in rejecting artificial correlations induced by common-distance biases. \label{sigmaofcc}}
\end{figure}

In practice, we implement the simulated dataset generation for a specific value of $c_L/c_z$ in two distinct ways, and we compare the results as shown in Fig. \ref{sigmaofcc}. First, we keep the redshift distribution fixed to a uniform distribution with lower limit $z_{\rm low} = 0$ and an upper limit $z_{\rm up} = 2$, and we draw the radio and gamma-ray luminosities from identical distributions with $\mu_0=0$ and a varying value of $\sigma_0$. In this way, we derive the black points in Fig. \ref{sigmaofcc}. Next, we keep the luminosity distributions fixed at $\mu_0=0$, $\sigma_0 =1$, and we draw redshifts from uniform distributions with varying upper and lower limits, always symmetric about $z=1$. In this way, we derive the red points in Fig. \ref{sigmaofcc}. 

For each dataset, we then evaluate the significance of the apparent correlation using the variation of our test that utilizes redshift binning; these results are shown with the  circles/solid lines in Fig.~\ref{sigmaofcc}. We compare these values with the significance estimate of Eq.~(\ref{student}) which does not account for any common-distance bias; these results are shown with the diamonds/dashed lines in Fig.~\ref{sigmaofcc}. For low values of $c_L/c_z$, the simple estimate of Eq.~(\ref{student}) returns false positives with high significance for these intrinsically uncorrelated datasets. Our method however correctly identifies these apparent correlations as artifacts of common-distance biases, and returns a significance value always consistent with no correlation. For higher values of $c_L/c_z$, common-distance biases are less important, and both significance estimates agree, returning a result consistent with no correlation. 

The roughly consistent, within noise, behavior of the black and red lines, despite the different method of implementation of the same value of $c_L/c_z$, implies that the $c_L/c_z$ ratio is a good way to quantify the way in which the dynamical ranges in the luminosity and redshift distributions induce common-distance biases in  correlations between different wavebands evaluated in flux space. As a rule of thumb, a value of $c_L/c_Z $ smaller than about 5 indicates that common-distance biases may be important, and the simple estimate of Eq.~(\ref{student}) (or, equivalently, permutation methods in flux space alone which do not account for the common distance modulation in each flux pair) should not be trusted as they might yield false positives.

\subsection{Demonstrations on simulated data: \\ Correlated datasets}

In the previous section we have shown that our proposed method successfully accounts for common-distance biases and returns results consistent with no correlation even when the simple estimate of Eq.~(\ref{student}) yields very significant false positives. Here we wish to examine whether this robustness against false positives comes at the expense of the power of the test in rejecting the null hypothesis of no correlation when the data are intrinsically correlated. In \S \ref{dataproduction2} we discuss how we generate intrinsically correlated datasets with minimal common-distance biases, and in \S \ref{deponN}  we discuss how the power of the test depends on the number of objects in the dataset, $N$, and on the apparent correlation strength, $r$, as well as how these dependencies compare with the simple formula of Eq.~\ref{student}. 

\subsubsection{Generation of correlated datasets}\label{dataproduction2}

To generate mock datasets with known intrinsic correlation signals, we assume that the radio and gamma-ray monochromatic luminosities at the frequencies of interest are linearly correlated\footnote{We adopt this assumption in the interest of simplicity for these demonstrations; this does not have to be the case in nature. Nonlinear correlations between the luminosities in the two wavebands will further complicate the relation between intrinsic and apparent correlation strength.}, with some scatter obeying a log-normal distribution: 
\begin{equation}
\log L _r = C + \log L_\gamma + \Delta \log L_r
\end{equation}
where $C$ is a normalization constant, and $\Delta \log L_r$ is normally distributed with mean $0$ and standard deviation $\sigma$, i.e. if $\Delta \log L_r =x$ then the probability density of $x$ is given by
\begin{equation}
p(x) = \frac{1}{\sqrt{2\pi} \sigma}\exp\left[-\frac{x^2}{2\sigma^2}\right]\,.
\end{equation}
Using Eq. \ref{kc}, this yields a relationship between radio and gamma-ray flux densities:
\beq\label{mockseq}
\frac{S_r}{S_{r,0}} = \frac{S_\gamma}{S_{\gamma,0}} (1+z)^{\alpha_r+\Gamma - 1} \times 10^{\Delta \log r}\,.
\eeq
The scatter in this intrinsic correlation is quantified by $\sigma$. We normalize the relation assuming that, for $z=\Delta \log r=0$, a 15 GHz radio flux density of $1 {\rm \, Jy}$ corresponds to a gamma-ray flux density of $10^{-8} {\rm GeV cm^{-2} s^{-1} GeV^{-1}}$ at 300 MeV. 

We generate mock datasets by starting from the set of 136 sources which (a) are detected by {\it Fermi} LAT at energies between 300 MeV and 1 GeV and are included in the First Fermi Catalog (1LAC, Abdo et al.~2010); (b) are included in the OVRO 40 M monitoring sample; (c) have known redshifts (see Ackermann et al.~2011 for the details of this sample). We use this set to obtain redshifts, gamma-ray fluxes, and gamma-ray spectral indices for our sources; for radio spectral indices, we use the historical values quoted in Ackermann et al.~2011.  We then use Eq.~(\ref{mockseq}) to obtain radio fluxes with a known correlation signal, by using the desired value of $\sigma$. 

The value of the $c_L/c_z$ ratio in the sample we use is $\sim 4$, so, according to the findings of \S \ref{robustness}, the effect of common-distance biases should be limited, and any apparent correlation between gamma-ray and radio emission should be primarily due to the intrinsic correlations we have imposed in the simulated datasets. In this case, we would expect a well-behaved test to return results that are close to the simple significance estimates of Eq.~(\ref{student}).

\begin{figure}
\includegraphics[width = 3in, clip]{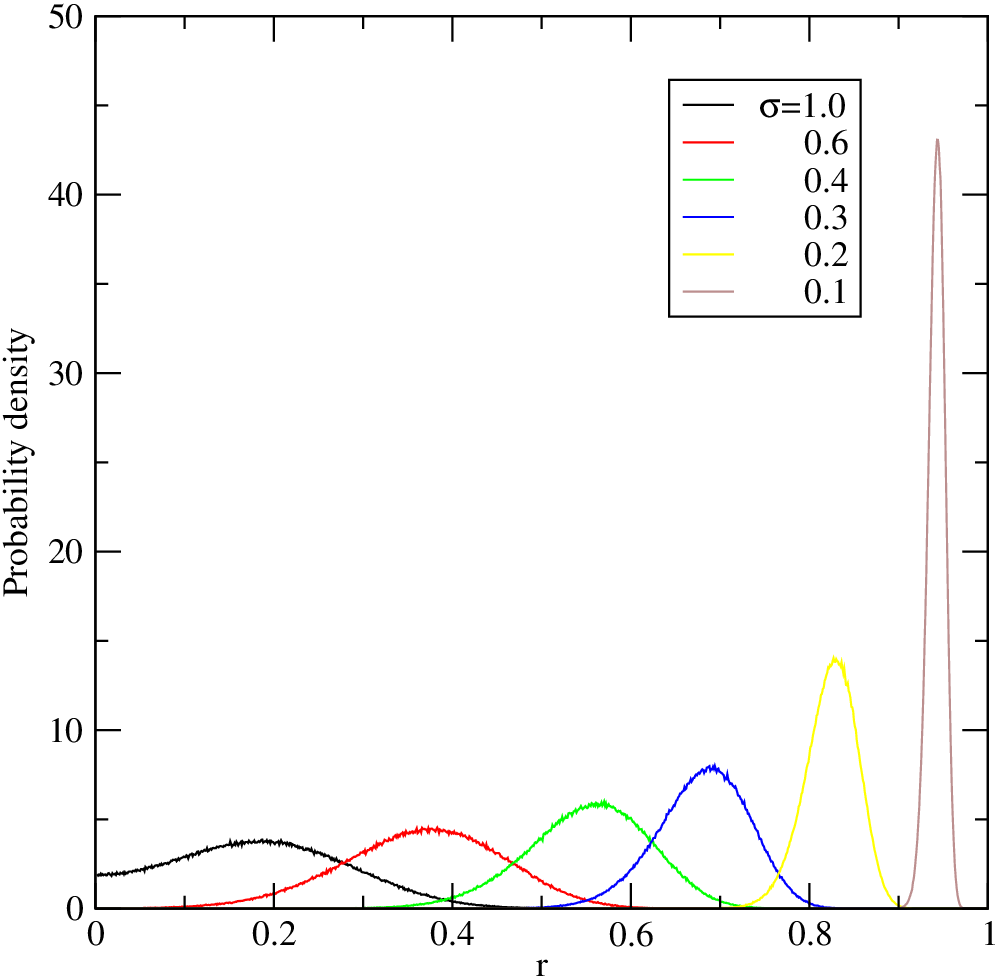}
\caption{Distribution of Pearson product-moment correlation coefficients $r$ that arise from random realizations of 80 objects obtained using 80 randomly chosen {\it Fermi} sources from 1LAC and Eq.~\ref{mockseq}, for various values of $\sigma$ as in legend.  \label{rofsigma}}
\end{figure}

Figure \ref{rofsigma} shows the distribution of Pearson product-moment correlation coefficients $r$ that arise from random realizations of 80 objects obtained in the manner described above,  for various values of the intrinsic correlation scatter $\sigma$.  The striking feature of this plot is that the distributions of possible $r$ values of ``observed'' flux/flux correlations arising from different random realizations of the same intrinsic luminosity/luminosity correlation sampled with the same number of points can be quite extended, with its width increasing with increasing $\sigma$. 

Even if we assumed that we knew the form of the underlying intrinsic luminosity/luminosity correlation, i.e. if, in our case, we assumed that Eq.~\ref{mockseq} holds exactly, and even with a relatively large sample (80 objects in this case), the observed value of the flux/flux correlation coefficient would only yield a rough and uncertain estimate of the scatter $\sigma$ of the underlying correlation, although the uncertainty of the estimate would improve for increasing values of $r$ (decreasing values of $\sigma$). 

\subsubsection{Dependence of significance on number of observations and apparent correlation strength}\label{deponN}

\begin{figure}
\includegraphics[width = 3in, clip]{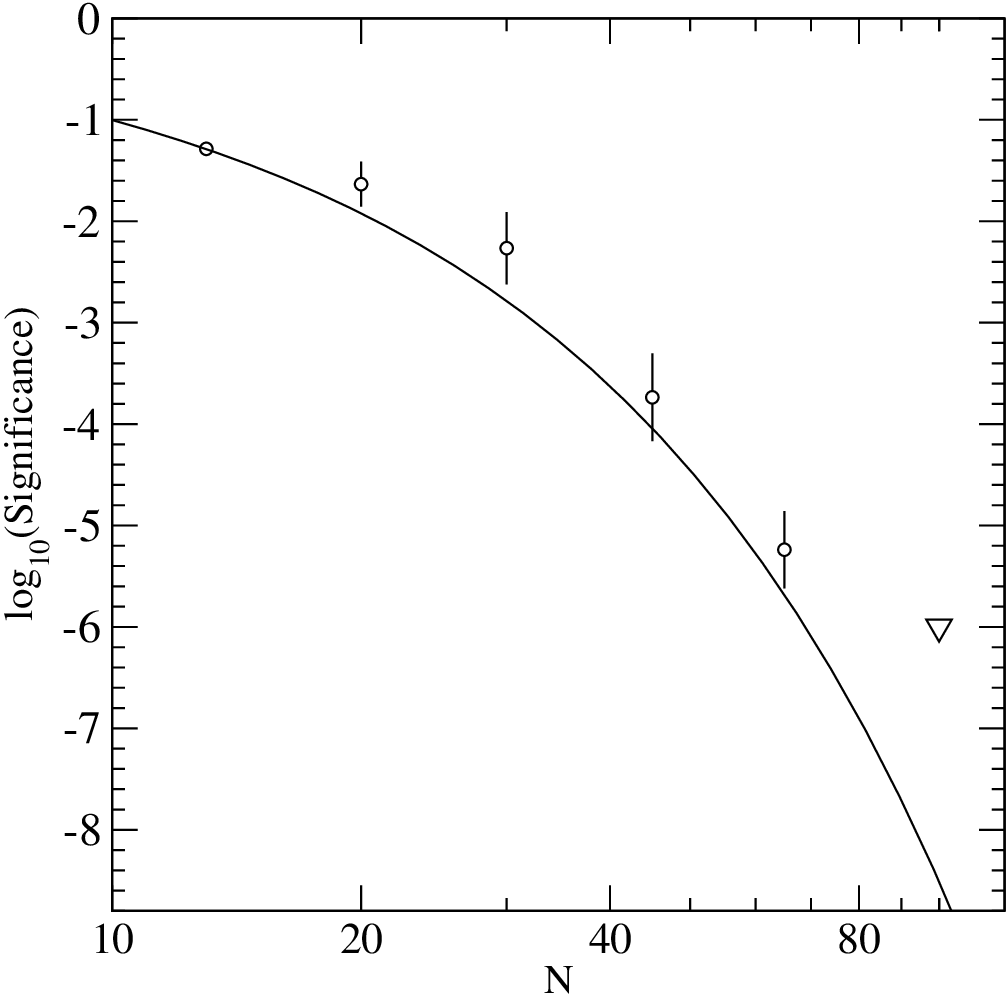}
\caption{Significance of an intrinsic correlation with $\sigma = 0.4$ sampled with $N$ objects and resulting in an apparent correlation strength of $r=0.55\pm0.001$, as a function of $N$. The points indicate the mean and error bars indicate the standard deviation of the calculated significances in 10 random implementations of the correlation. The downwards triangle indicates an upper limit for $N$=100, where the probability of the correlation to arise by chance was always found to be $<10^{-6}$ (none out of $10^7$ scrambled datasets had an $|r|$ at least as big as the ``data'').  The solid line shows the result of Eq.~(\ref{student}) for $r=0.55$ and varying $N$.\label{sofn}}
\end{figure}

Figure \ref{sofn} shows the dependence of the calculated significance of a correlation of fixed apparent and intrinsic strength (i.e, fixed values of $r$ and $\sigma$) on the number of objects in the sample. In the example presented here, mock datasets of  $N$ objects were generated as described in \S \ref{dataproduction2}, using a fixed 
 intrinsic correlation scatter  $\sigma = 0.4$, and requiring an apparent correlation strength of $r=0.55\pm0.001$. For every value of $N$ plotted in  Fig.~\ref{sofn}, 10 such mock datasets were produced, and for each dataset the significance was evaluated using the redshift-bin-splitting variation of the test (with the number of redshift bins chosen, for each value of $N$, as discussed in \S \ref{thesplit}.)
 The datapoints in Fig.~\ref{sofn} represent the the mean of $\log_{10}$(Significance) for these 10 realizations, and the error bars indicate the standard deviation of the 10 values of $\log_{10}$(Significance). The solid line shows the result of Eq.~(\ref{student}) for $r=0.55$ and varying $N$. Even this modest correlation with appreciable scatter can be established at high significance (better than $\sim 10^{-5}$) with 60 or more objects.

\begin{figure}
\includegraphics[width = 3in, clip]{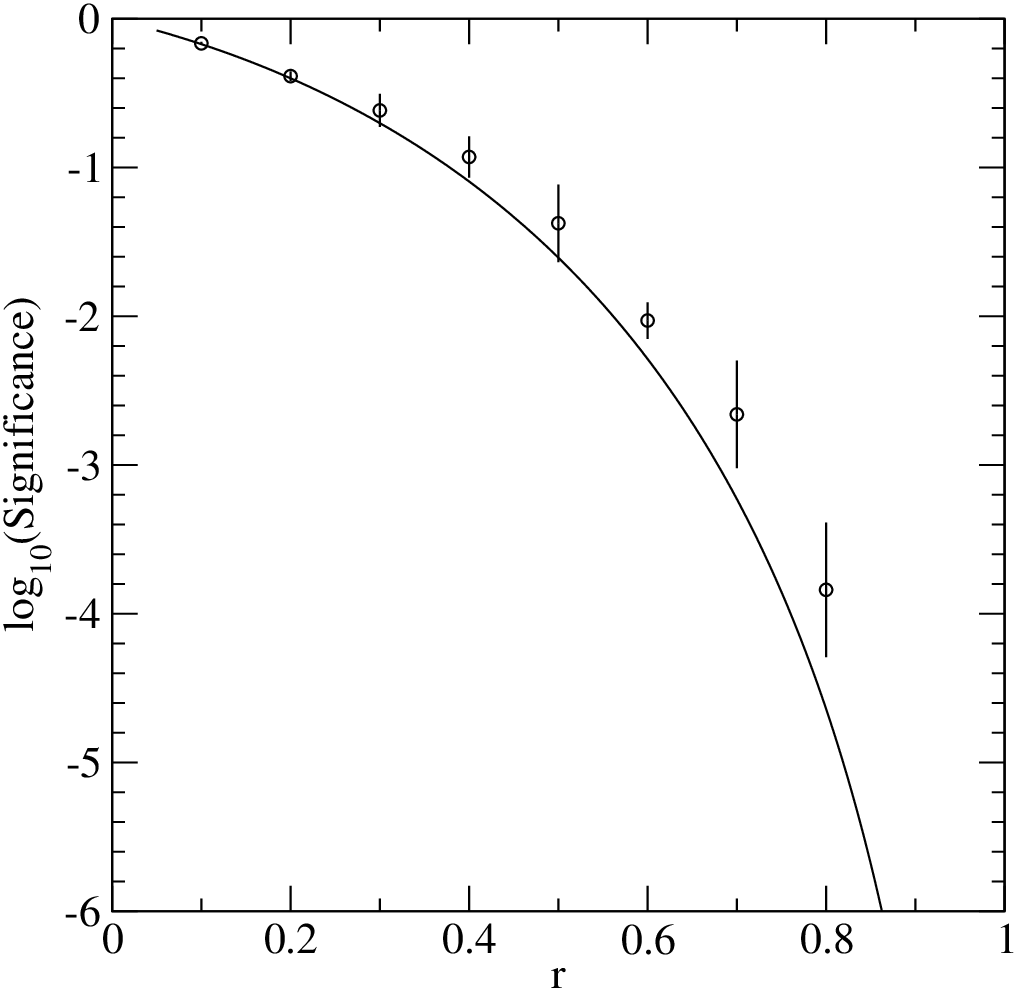}
\caption{Significance of an intrinsic correlation with $\sigma = 0.6$ sampled with 20 objects and resulting in an apparent correlation strength of varying $r$, as a function of $r$. The points indicate the mean and error bars indicate the 1$\sigma$ variation of the calculated significances in 10 random implementations of the correlation.  The solid line shows the result of Eq.~(\ref{student}) for $N=20$ and varying $r$. \label{sofr}}
\end{figure}

Figure \ref{sofr} shows the dependence of the significance that the redshift-bin-splitting variation of our method yields as a function of the apparent correlation strength (as quantified by $r$), when the underlying, intrinsic correlation and the number of objects are fixed. We have used an intrinsic correlation with relatively large scatter ($\sigma =0.6$), sampled with a relatively small number of objects ($N=20$). As it is obvious from Fig.~\ref{rofsigma} (red line), this large scatter can result in a variety of apparent correlation values. Again, for each value of $r$ plotted in Fig.~\ref{sofr}, we have generated 10 mock datasets as described in \S \ref{dataproduction2}, demanding that their apparent correlation strength is within $0.001$ of the plotted $r$ value. For each dataset the significance was evaluated using the redshift-bin-splitting variation of the test (with the number of redshift bins chosen, for each value of $N$, as discussed in \S \ref{thesplit}.)
 The datapoints in Fig.~\ref{sofr} represent the the mean of $\log_{10}$(Significance) for these 10 realizations, and the error bars indicate the standard deviation of the 10 values of $\log_{10}$(Significance). The solid line shows the result of Eq.~(\ref{student}) for $N=20$ and varying $r$.

As we can see in both Fig.~\ref{sofn} and Fig.~\ref{sofr}, the significance returned by our method in the absence of strong common-distance biases and for correlated datasets is consistent with, or only marginally worse than the results of the simple estimate of Eq.~(\ref{student}). We therefore conclude that the robustness of our method against common-distance biases in uncorrelated datasets does not come at the expense of the power of the method in the case of correlated datasets.   

\section{Discussion and Conclusions}

In this paper, we have introduced a data-randomization method for assessing the significance of apparent correlations between radio and gamma-ray emission in blazar jets, accounting explicitly for biases introduced through a common distance, small sample size, and complex or subjective sample selection criteria. Our method is designed to be conservative and applicable even to small samples selected with subjective criteria. 

 An application of this technique to the  first {\it Fermi} catalog of point sources (Abdo et al.\ 2010) has been discussed in Ackermann et al.~(2011), which also discusses the dependency of the strength and significance of the radio/gamma flux correlation on gamma-ray photon energy, and on the concurrency of the two datasets. A study of the dependency of correlation strength and significance on radio frequency can be assessed by application of this method to multi-frequency radio monitoring data, such as the results  of the F-GAMMA Program (Angelakis et al.~2010; Fuhrmann et al.\ 2012, in preparation).

Using simulated datasets of intrinsically uncorrelated data, we have demonstrated that our proposed method is robust against artificial correlations induced by common-distance biases, and returns results consistent with no correlation even when simple face-value significance estimates that do not account for these biases would have incorrectly claimed highly significant correlations. We have shown that the effect of these biases can be quantified by the ratio of the coefficients of variation of the luminosity distribution (in the waveband which has the widest luminosity distribution) over the redshift distribution.  When this ratio is lower than $\sim 5$ false positives are possible when the biases are not accounted for, with their significance increasing with decreasing value of the ratio. In addition, using simulated datasets of intrinsically correlated data, we have shown that our method can establish existing correlations with significance comparable to that of other tests, and thus its robustness against false positives does not come at the expense of its power in rejecting the null hypothesis. 

As our method is designed to be applied to astronomical datasets, we have implemented it in such a way to directly address the limited flux dynamical range that is generally encountered in such data. Astronomical datasets are generally expected to have a low-flux limit in each frequency due to the limited sensitivity of any given observing instrument, and a high-flux limit corresponding to the most favorable combination of luminosity/distance that happened to occur given our position in the universe, which is generally determined by chance, and fixed by the observed dataset. Simulated data with one or two of the fluxes outside these limits  represent situations possible in nature but impossible to observe in the specific experiment. Such simulated data coud result in apparent correlations in our simulated samples induced by a single very bright or very faint object much brighter or fainter than the objects in our actual sample, while most of the other pairs would be scattered in a limited area of the flux/flux space (the classical case of an artificial correlation seen when an uncorrelated scattered cluster of points is combined with a single point far away from the main cluster). Such configurations would be impossible in our actual observed datasets, but could be frequent in our simulated datasets, thus biasing the simulated datasets toward much higher correlation coefficients, and artificially reducing the significance of any correlation seen in the observed datasets. We exercise care to avoid this bias by limiting the flux dynamical range of our randomized data to that of the observed sample and by rejecting simulated flux pairs outside this range. 

We caution the reader that our test does not in any way account for the effects of non-simultaneity.  Ideally the test should be applied to data in different frequencies averaged over the same time interval. The spectral indices used in each band to implement the K-correction should also be concurrently measured with the flux averages. Such concurrent spectral indices are straight-forward to obtain in gamma rays, however radio monitoring is routinely performed at a single waveband (as is the case for the OVRO 40 M Monitoring Program), and in practice only archival radio spectral indices are available. It is thus fortunate that our test is robust against small changes in the value of the radio spectral index used in the K-correction. This property of the test can be understood by taking into account that blazars are spectrally flat at radio frequencies so the effect of the K-correction in radio is small to begin with. We have confirmed this by alternatively using archival radio spectral indices measured for each source, or a uniform value of $\alpha_r=-0.5$ across all sources; the evaluated significance in the two cases did not change appreciably (fractional change in the quoted significance less than $0.01$). This result can be explicitly confirmed by using multi-frequency simultaneous data from the F-GAMMA Program where simultaneous radio spectral indices can  be obtained. These tests are described in detail in Fuhrmann et al.\ 2012 (in preparation). 

In contrast, the test is quite sensitive to the redshift of the sources included in the sample under consideration, as shown in Ackermann et al.~(2011): as the main purpose of the test is to assess the effect of distance biases, the calculations involved are sensitively dependent on said distances. For this reason the test should only be used on samples with known redshifts for all members. 

Finally, we stress that this test in itself does not assess the strength of the intrinsic correlation between flux densities of different frequencies. It only addresses its statistical significance, i.e. the probability that an apparent correlation as strong or stronger than the observed one can be obtained from intrinsically uncorrelated data due to observational biases. A correlation may be very weak but picked up at high significance if the dataset is large and the data quality is high; in contrast, a strong correlation in a very small sample may not be very significant. It is similarly important when the test returns a low statistical significance to carefully distinguish between lack of evidence for correlation and evidence for intrinsically uncorrelated data. Intrinsic lack of correlation cannot generally be established. However it is possible to show that a correlation with strength above a certain threshold would have been picked up at a given level of significance by a particular test.

\acknowledgements{We thank Andy Strong and the anonymous referees for insightful comments which improved this manuscript. The OVRO 40 M program is supported in part by NASA grants NNX08AW31G  and NNG06GG1G, and NSF grant AST-0808050. Support from the Max-Planck Institut f\"{ur} Radioastronomie for upgrading the OVRO 40 M telescope receiver is also acknowledged. We are grateful to Russ Keeney for his tireless efforts in support of observations at the Owens Valley Radio Observatory. VP acknowledges support for this work provided by NASA through Einstein Postdoctoral Fellowship grant number PF8-90060 awarded by the Chandra X-ray Center, which is operated by the Smithsonian Astrophysical Observatory for NASA under contract NAS8-03060, and thanks the Department of Physics at the University of Crete for their hospitality during the completion of part of this work. 
 WM acknowledges support from the U.S. Department of State and the Comisi\'{o}n Nacional de Investigaci\'{o}n Cient\'{i}fica y Tecnol\'{o}gica (CONICYT) in 
Chile for a Fulbright-CONICYT scholarship. 
}

\end{document}